# Workflow-Based Big Data Analytics in The Cloud Environment:

## Present Research Status and Future Prospects

Samiya Khan
Student Member, IEEE
Department of Computer Science,
Jamia Millia Islamia
New Delhi, India
samiyashaukat@yahoo.com

Kashish A. Shakil
Student Member, IEEE
Department of Computer Science,
Jamia Millia Islamia
New Delhi, India
shakilkashish@yahoo.co.in

Mansaf Alam
Member, IEEE
Department of Computer Science,
Jamia Millia Islamia
New Delhi, India
malam2@jmi.ac.in

*Abstract*— Workflow is a common term used to describe a systematic breakdown of tasks that need to be performed to solve a problem. This concept has found best use in scientific and business applications for streamlining and improving the performance of the underlying processes targeted towards achieving an outcome. The growing complexity of big data analytical problems has invited the use of scientific workflows for performing complex tasks for specific domain applications. This research investigates the efficacy of workflow-based big data analytics in the cloud environment, giving insights on the research already performed in the area and possible future research directions in the field.

## I. INTRODUCTION

Scientific and business analytics entails several applications that require the use of scientific workflows to mitigate the complexities involved. In fact, in fields like Astronomy, Social Science, Bioinformatics and Neurosciences, in addition to several others, scientific workflow management systems are found to be effective, so much so that they are irreplaceable in their realm of usage [17, 18, 19].

Similar to the traditional data management systems and computing infrastructures that have been proven insufficient for the challenges posed by big data analytics; similarly, traditional scientific workflows are also unable to deal with the growing scale and computational complexity of big data analytics tasks. It is at this point that Cloud Computing seems like an appropriate solution to the big data problem.

However, devising a synergistic model that can help leverage the power of traditional scientific workflows and use it with cloud computing to solve big data problems is a challenge [4]. There are several ways and approaches that have been explored in this direction. This research paper explores the different approaches and the systems that have been implemented to put this idea into practice. In addition, this paper will also explore the possibility of future research.

The organization of the rest of this research paper has been described below. The first section introduces scientific workflows, giving insights into fundamental definitions and concepts related to the topic. As part of this section, a brief comparison of the different scientific management systems and frameworks that support cloud-based execution has also been presented.

The next section throws light on how and where the cloud paradigm fit into the scientific workflow concept for big data analytics. The section that follows elaborates on the challenges and issues that arise in bringing together this synergistic approach. The last section of the paper explains existing systems and research gaps that exist and can be worked upon. The paper concludes with a remark on the future research direction in the usage of scientific workflows for cloud-based big data analytics.

## II. UNDERSTANDING SCIENTIFIC WORKFLOW CONCEPTS

At the basic level, a workflow can be explained as a logical sequence of activities or data processing tasks, which works on the basis of predefined rules. The fundamental usage of a workflow is to automate any process. Typically, there are two types of workflows - (1) scientific workflows and (2) business workflows. Scientific workflows find applications in the field of scientific computing for automating scientific experiments and processes [7] while the latter is used for automating business processes.

There are several ways in which scientific workflows are represented. The most commonly used representations are Directed Acrylic Graphs (DAG) and Directed Cyclic Graphs (DCG) [4]. The two fundamental entities that need to be described with respect to a scientific workflow are activities and tasks. An activity is a logical step that needs to be performed as part of a scientific workflow [8]. On the other hand, a task is an instance of an activity [9]. Therefore, a task represents the execution of an activity, which happens only once.

The transition of a workflow from its initiation to its completion is termed as the scientific workflow lifecycle. Moreover, the Scientific Workflow Management System (SWfMS) performs initiation and management of workflow execution. Gorlach et al. [11] proposed that a scientific workflow is divided into four phases namely deployment phase, modeling phase, execution phase and monitoring phase.

There are several scientific workflows that are used as solutions to field-specific and generic problems. Liu et al. [4] compared Swift [26], Pegasus [32], Taverna [35], [31], Galaxy [34], Chiron [36], Askalon [33], Triana [1] and WS-PGRADE/gUSE [6], of which the first eight are typical Scientific Workflow Management Systems and WS-PGRADE/gUSE is a gateway framework. Systems and frameworks discussed in this paper support dynamic scheduling, independent parallelism and cloud-based scientific workflow execution. A comparison of these systems and frameworks has been presented in Table 1.

TABLE I. COMPARISON OF SCIENTIFIC WORKFLOW MANAGEMENT SYSTEMS AND WORKFLOWS

| SWfMS/Feature | Workflow Structure | Static Scheduling | Information Sharing | User Interface | Special Feature | Applications |
|---|---|---|---|---|---|---|
| WS-PGRADE/gUSE | DAG | ✓ | ✓ | GUI | Allows execution of scientific workflows in Distributed Computing Infrastructures | Used for executing workflows for astronomy, biology, seismology and neuroscience. |
| Pegasus | DAG | ✓ | ✗ | Textual | High Performance and Scalable | Used for executing workflows for astronomy, biology etc. |
| Kepler | DAG and DCG | ✓ | ✗ | GUI | Desktop-based GUI | Used for executing workflows for astronomy, biology etc. |
| Chiron | DAG and DCG | ✗ | ✗ | Textual | Workflow Parallelization based on algebraic approach | Large scale scientific experiments |
| Askalon | DAG and DCG | ✗ | ✓ | GUI (Desktop and Web) | Allows execution of scientific workflows on a multi-site cloud environment | Scientific applications |
| Galaxy | DAG and DCG | ✗ | ✓ | GUI (Accessible from web) | Web-based system for genomic research | Used for executing Bioinformatics workflows only |
| Swift | DAG and DCG | ✗ | ✗ | Textual | High Performance and Scalable | Used for executing workflows for astronomy, biology etc. |
| Taverna | DAG | ✗ | ✓ | GUI | Desktop-based GUI | Used for executing workflows for astronomy, biology etc. |
| Triana | DAG and DCG | ✗ | ✗ | GUI | Can use P2P Services | E-Science applications |

## III. CLOUD FOR SOLVING TRADITIONAL SCIENTIFIC WORKFLOW IMPLEMENTATION ISSUES

Data is being generated in this world at an alarming rate in view of the ever-increasing popularity of social networks like Facebook, eBay and Google+, amongst many others. Most of the data being generated is already on the cloud environment. In fact, according to an estimate given by Schouten [12], 50% of the total data will be cloud-based by the year 2016. The fundamental challenge in the management of this data is its storage and processing, as the present-day systems cannot support the same.

In addition to the above-mentioned, processing of data makes use of complex, computing intensive algorithms. This requires systems that can handle the processing requirement of data mining algorithms, making high performance computing the second requirement of big data analytics that any system claiming to be an effective solution needs to fulfill.

The Cloud Computing environment is an apparent solution to the big data problem [30]. Firstly, the cloud provides an operative storage solution for the huge data storage problem. The cloud adopts pay-as-you-go model, which is primarily why it is able to offer a cost-effective solution to the problem. Besides this, the concept of commodity hardware introduced by cloud computing allows user to get the hardware required without the need to buy it, giving a scalable solution to the computing hardware requirement.

These are the reasons why most enterprises and the scientific community have chosen to adopt the cloud for big data management and analysis. The availability of effective, efficient and open-source ecosystems like Hadoop has facilitated the adoption process immensely. The complexity of the problem further intensifies when cloud-based big data analytics requires the execution of scientific workflows, which are data intensive in nature.

There are three main facets of this increasing complexity. Firstly, the types and sources of data are diverse. Secondly, the whole concept of using distributed computing for data processing is based on moving code to the data. However, this is not always possible because of compatibility issues and proprietorship of the code. Lastly, it is not possible to keep all the data throughout the lifecycle of the execution of a workflow. Therefore, redundant data needs to be removed. Several workflows have been proposed for running data intensive workflows, which shall be discussed in the following sections.

Evidently, executing data intensive workflows shall require handling of large datasets. Therefore, the most obvious solution to ease the execution process is to use parallelization. Cloud is a great solution to solve the need for unlimited resources in data intensive workflow execution. Performance and cost optimization efforts in this direction are focused towards improving the scheduling algorithm [21]. Parallelization may be implemented at the single site cloud-level or a multi-site cloud-level. Multi-site cloud parallelization is an area that has gained immense research interest lately. Moreover, the use of workflow partitioning techniques for efficient multi-site cloud parallelization is also being explored.

## IV. Scientific Workflows On The Cloud: Challenges/Issues Faced

In the scientific workflow context, traditionally, they are implemented on grids [10] or clusters, workstations and supercomputers. This implementation faces several limitations and obstacles, the most profound of which are scalability and computing complexity. Apart from these, several other issues like resource provisioning have also been known [3]. This section discusses some of the most prominent challenges facing the development and use of scientific workflows for big data analytics in the cloud environment.

### A. Increasing Data Size and Computing Complexity

The input that goes into scientific workflows and the output that comes out of the same are data objects that are distributed in nature. The type, size and complexity of these data objects shall vary. The data coming from all the different sources of scientific computing like sensors, experiments and networks is increasing at a rapid rate and giving rise to the immensely advocated concept of 'data deluge' [13].

According to this concept, the whole theory of scientific research is changing. While, earlier, a question was asked and data was collected that can answer the question, today, we have heaps of data and retrospectively, what we are looking for is the question that this data heap can answer. Evidently, the scale and size of data is huge. In order to handle the growing complexity of data storage and processing, several thousands of computation nodes may be used. This will not only make the operation feasible, but it shall also make it efficient.

### B. Provisioning of Resources

The process of allocating resources to a scientific workflow to operate in terms of storage and computing resources is termed as resource provisioning. Typically, a scientific workflow is allocated resources as and when the scientific workflow is deployed. Moreover, these resources are fixed prior execution. Evidently, the scale of the scientific workflow is limited by resource allocation. At another level, the scale of scientific workflow is also limited by the total resource pool size. In order to allow smooth and scalable execution of scientific workflows, an efficient dynamic resource provisioning shall be required.

There are several other challenges in this category of research. One of the fundamental challenges that need to be clearly addressed in this context is how a resource can be best represented for scientific workflows [4]. In addition, every workflow has a set of supported tools and resource types. Identifying whether a resource type or tool is compatible with a workflow is also a daunting research task. Recent research in this field has been focused towards automated provisioning and developing algorithms for effective resource provisioning. One such system designed for the cloud to manage provisioning of resources for the workflow is the wrangler system [2].

### C. Getting Heterogeneous Environments to Work Together

Research has always been a collaborative work and as the world is converging after the advent of Internet, geographical boundaries and distances are no longer a limitation. However, scientific projects running at varied geographical locations add to the complexity of using workflow management systems in view of the fact that the execution environments used by different organizations may be different.

Therefore, the interactions between the execution environment on the host system and the workflow management system need to be smoothened for resource management, security and load balance, besides others [16]. When it comes to heterogeneous execution of scientific workflows, the varied computational ability and performance of systems may also impact the traditional scientific workflow execution.

## V. Scientific Workflows In The Cloud

Workflows are described as complex graphs, unfolding the concurrent tasks that any concerned application may include. Evidently, any data analytics task will require data access, processing and visualization. Therefore, workflow tasks must address these components of the system effectively and efficiently.

An obvious approach for using workflows in the cloud environment is to refactor the existing scientific workflows according to the Cloud Computing paradigm. One of the first works that proved the viability and effectiveness of the cloud for running scientific workflows was Keahey and Freeman [14]. As part of this research, a Nimbus Cloudkit was developed, which was made available to the scientific community for satisfying their infrastructure and resource needs.

Vöckler et al. [22] implemented a cloud-based scientific workflow application for processing astronomical data. The Pegasus workflow management system was deployed on multiple and different clouds to test the viability of the system for the application stated. The findings of this experiment included a conclusion that the user experience was not affected by the underlying differences in the different clouds used and users were able to accomplish basic tasks with ease, indicating that the management overheads of the system were ignorable.

Although, this seems like an obvious option, it is impractical considering the high complexity of scientific workflows. Developers will have to invest a huge amount of effort and time to implement the application logic and mitigate the challenges of integrating the workflow logic with the underlying cloud.

A more feasible approach is the integration of Scientific Workflow Management Systems into the cloud environment. In this way, traditional workflows will not have to be refactored, addressing the challenges associated with the same, and they can still be used to process cloud data. This concept has given rise to Cloud Workflow platform, which is provisioned to the users as a service.

There are several advantages of using this approach. Some of the fundamental benefits include scalability, flexible resource allocation, easier deployment of applications and a better return on investment from the organization's point of view. However, what this approach also adds to this list is an increased overhead, but this facet can be ignored keeping in mind the multifold benefits of scientific workflow management systems that this approach allows researchers to leverage.

Typically, workflow-based data mining on the cloud makes use of the service-oriented approach. According to Talia [15], there are three advantages of using this approach namely, execution scalability, distributed task interoperability and a flexible programming model. The scalable model of execution provided by this approach helps in considerably reducing the completion time of the task. There are many frameworks and architectures that have been proposed. They have been discussed below.

Lin et al. [5] proposed one of the first architectures given for Scientific Workflow Management systems. It introduced a four-layer architecture with workflow management layer, presentation layer, operational layer and task management layer, being the four layers, considering visualization, analytical engine and data acquisition. However, this architecture was a base architecture that had no provisions for management of security issues. Fig. 1 illustrates the architecture and the components that it includes.

Zhao et al. [16] proposed a service framework for integration of Eucalyptus and OpenNebula with Swift Workflow Management System that addresses these challenges. Fig. 2 illustrates the basic layout of this framework. They have also presented an implementation based on this service framework and makes use of the Montage Image Mosaic Workflow.

However, porting of other traditional workflows like VIEW and Taverna is yet to be explored in addition to investigating the possibility of automatic deployment of workflow applications in the virtual cluster. This framework provides the background for designing efficient scientific workflow management systems.

Fig. 3 shows the functional architecture of SWfMS given by Liu et al. [4]. The user services layer caters for user functionality while the presentation layer is the GUI that shall be presented to the user for giving instructions to the system.

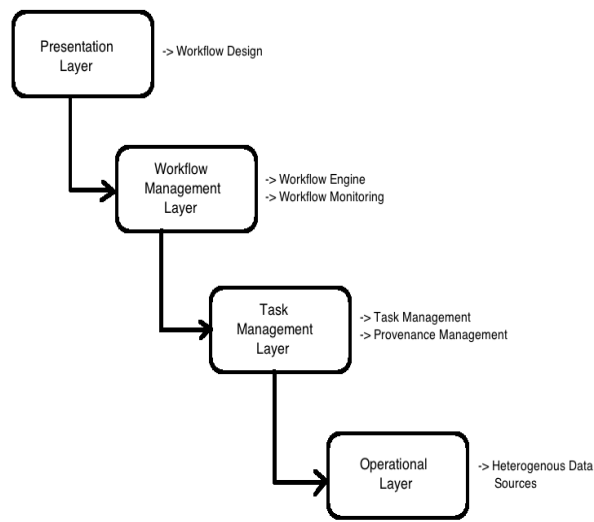

Fig. 1. Reference Architecture by Lin et al. [5]

The WEP generation layer generates the workflow execution plan (WEP), which is used by the system to interpret the user instructions and determine the flow of execution of the workflow. The generated WEP is given to the WEP execution layer for execution. The last layer is the infrastructure layer. It delivers the required computing and storage resources.

Li, Song and Huang [21] proposed scientific workflow management system architecture for manufacturing big data analytics. The architecture is based on the three architectures previously mentioned and divides the system into four layers namely, infrastructure, management, service and application layers. This system is implemented as a secondary system over Kepler, an existing system.

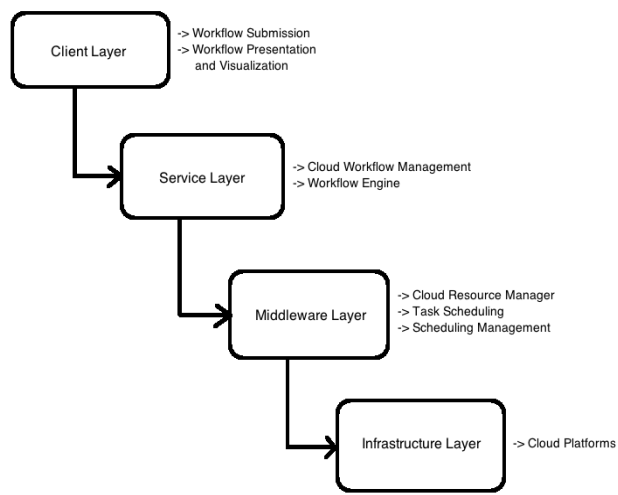

Fig. 2. The Service Framework by Zhao et al. [16]

The key issue in management systems is to devise an efficient scheduling algorithm. The objective of a scheduling algorithm is mapping of tasks and resources. In the present context, this should be distributed and heterogeneous in nature. Therefore, there is a need for deterministic workflows that require task parameters and resource configuration/availability. However, the workflow must not need an input on where these resources are located [24].

In view of this, workflows that require an abstract mathematical model as input are chosen. The workflow is represented in the form of a DAG. Here, nodes represent tasks edges are indicative of the relationships between tasks [27]. Several heuristics and meta-heuristics algorithms are present to solve this NP Complete problem keeping in mind the resource QoS and execution cost [25, 28, 29, 30].

Execution of scientific workflows can be performed on the Hadoop YARN. Hi-WAY [42] gives an engine for scientific workflow execution. In other words, it provides an application master that can control scientific workflow task execution on top of YARN. However, some fundamental shortcomings exist in this engine. One such shortcoming comes from the fact that the containers allocated for task execution are uniform for all applications and tasks, which leaves room for optimization. Therefore, customized container allocation can considerably improve the performance of the engine.

This research paper proposes an MP (Max Percentages) algorithm, which resembles Min-Min, Sufferage and Max-Min algorithms and was tested to outperform classic algorithms both in terms of load balancing and time of completion. This algorithm is based on the intrinsic correlation between resources and tasks for performance optimization.

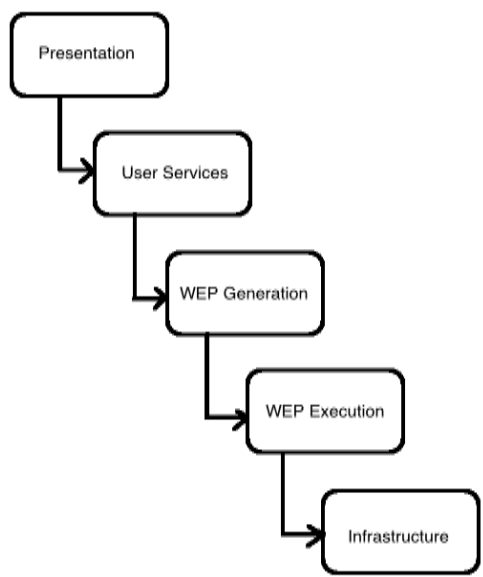

Fig. 3. SWfMS Architecture by Liu et al. [4]

However, security and performance of the system for dynamic heterogeneous environments have not been tested and can be explored. This research can be taken as reference for creating more efficient scientific workflow architectures and scheduling algorithms for generic applications.

## VI. SWFMS IN MULTISITE CLOUD

Most of the research work being performed in this field is related to scientific workflow management systems' deployment in the cloud environment and their execution in the multisite cloud environment. A recent work in this area provides architecture for efficient execution of scientific workflow management systems using the distributed approach [40].

The described architecture has shown significant cost reduction and can be treated a good base architecture, which can be improved using dynamic scheduling, distributed provenance management, use of multisite spark and better data transfer techniques. The master-slave architecture can be changed to peer-to-peer architecture for improving the system. A significant attempt in this direction has incorporated multi-objective scheduling in the architecture. It has been proposed that location awareness can bring about a noteworthy improvement in data transfer issues and performance [39].

Existing literature indicates research work on provenance management, metadata management [37] and big data management [38] in the multisite environment. In order to effortlessly amalgamate the SWfMSs in the cloud environment, it is important to include cloud configuration parameters and resource descriptions to provenance data. Ahmad [41] describes how execution reproducibility is affected in the light of cloud-aware provenance.

Effective management of metadata and big data generated in the SWfMS can bring considerable improvements in the performance of the system. It is shown that the use of distributed approach for metadata management betters the system performance by as much as 50% [39]. This approach can be extended for heterogeneous multisite environment.

Evidently, the data distributed across geographical locations suffers from management issues like cost-related tradeoffs, low latency and high throughput. Challenges are all the more magnified considering the volume of big data. An implementation on Azure cloud presents a uniform data management system. This system allows spreading of data geographically separated locations. However, the existing system uses per-site registration of metadata. It is proposed that a hierarchical system of global nature must replace the existing metadata registration system.

While working in the Multisite Cloud environment, it is important to mention that the included clouds may be heterogeneous in nature and the individual cloud providers are yet to provide interoperability. To explore the management and deployment of workflows over heterogeneous clouds, a broker-based framework was proposed by Jrad, Tao and Streit

[20], which allows automatic selection of target clouds. Kozlovsky et al. [23] provided an internal architecture to enable compatibility for workflow management systems by resolving the DCI interoperability issues prevalent at the middleware level.

Some other research in the area has been targeted towards creating specific applications for domains like astronomy [19], geo-data analysis [18] and Bioinformatics [17], in addition to several others. Talia [15] demonstrated the use of Data Mining Cloud framework and established the effectiveness and linear scalability of this framework in bioinformatics, network intrusion and genomics. However, this framework was not tested for big data and complex data mining.

## VII. CONCLUSION

Scientific Workflow Management Systems use a viable approach for integrating scientific workflows with cloud-based big data analytics. Some of the main advantages of this approach are better scalability, higher flexibility and easier deployment. In view of these benefits, several frameworks, architectures and scheduling algorithms have been proposed.

However, research on scientific workflow management systems is still in its infancy. The paper compares the different frameworks and architectures proposed and in use for exploiting scientific workflows in the cloud environment. It explores the possibility of using the same for big data analytics. While many systems have been designed and implemented, security issues remain unaddressed in all these solutions. Moreover, optimization of scheduling algorithms and multi-site cloud execution of workflows need attention in future research related to this field.

## ACKNOWLEDGMENT

This work was supported by a grant from "Young Faculty Research Fellowship" under Visvesvaraya PhD Scheme for Electronics and IT, Department of Electronics & Information Technology (DeitY), Ministry of Communications & IT, Government of India.